\documentclass[proof]{WileyASNA-v1}

\articletype{Article Type}%

\received{}
\revised{}
\accepted{}

\begin{document}

\title{Barium abundances of A--F--G type stars in the Hyades cluster}

\author[]{Yoichi Takeda}

\authormark{Y. TAKEDA}

\address[]{ 
\orgaddress{\state{11-2 Enomachi, Naka-ku, Hiroshima-shi, 730-0851}, \country{Japan}}}

\corres{\email{ytakeda@js2.so-net.ne.jp}}


\abstract{
With an aim of clarifying the extent and parameter-dependence of 
compositional anomaly of barium in A-type stars, Ba abundances were
spectroscopically determined based on Ba~{\sc ii} 6141/6496 lines for 
89 (23 A-type and 66 F--G-type) main-sequence stars belonging to 
the members of Hyades cluster by taking into account the non-LTE effect 
and the hyper-fine-structure effect. While the non-LTE effect tends to 
strengthen lines in G stars, it acts in the direction of line weakening 
in the regime of A stars due to increasing importance of overionization.
The Ba abundances of G stars turned out almost constant ($\langle A \rangle = 2.33$),
indicating that the primordial composition of Ba in Hyades is mildly 
supersolar by $\simeq +0.2$~dex. 
In contrast, A-type stars show Ba overabundances of considerably 
large dispersion (0~$\lesssim$~[Ba/H]~$\lesssim 2$). 

Since this Ba excess 
tends to increase with an increase/decrease in $T_{\rm eff}$/$v_{\rm e}\sin i$,
these two parameters may be essential for producing or controlling the anomaly.
Regarding Hyades F-type stars, their Ba abundances are not uniform but 
show a broad depression (by $\lesssim 0.3$~dex) around $T_{\rm eff} \sim 6500$~K, 
interestingly coinciding with the location of Li-dip.
}


\maketitle

\section{Introduction}

It is known that an appreciable fraction of non-magnetic A-type stars 
(especially  comparatively slow rotators) in the effective temperature 
range of $T_{\rm eff} \sim$~7000--10000~K show characteristic abundance 
peculiarities often called ``Am phenomenon''.\footnote{
Although this term originally stems from the abundance characteristics of classical
A-type metallic-line (Am) stars, it is often used irrespective of whether a star is 
classified as Am or normal in its spectral type, because various degree of chemical anomaly 
actually exists without any clear-cut distinction between peculiar and non-peculiar stars.}
That is, while specific comparatively lighter elements (C, N, O, Sc, Ca) are apt to be 
deficient, heavier elements (such as Fe group, s-process, rare earths) tend to be 
overabundant (see, e.g., Fig.~5 in Smith 1996).    

Among these, barium (Ba; atomic number $Z = 56$) is one of the important elements 
exhibiting abundance anomaly of especially large extent (up to $\sim 2$~dex or even more).
Because of this remarkable feature, this element can be exploited for detecting 
peculiar A and F stars from large-scale surveys based on low-resolution spectrograms
(Xiang et al. 2000).

However, our understanding on the quantitative nature of Ba anomaly in A-type stars 
is still insufficient, especially regarding the zero-point of abundance reference. 
Takeda et al. (2008) carried out abundance determinations for 46 field A-type stars 
to investigate how chemical peculiarities depend upon rotational velocities 
($v_{\rm e}\sin i$), and found a tight correlation between [Ba/H]\footnote{
As usual, [X/H] is the differential abundance of element X (of a star) 
relative to the solar abundance; i.e.,[X/H] $\equiv A_{*}$(X) $- A_{\odot}$(X). 
} and [Fe/H] as well as a systematic 
$v_{\rm e}\sin i$-dependent trend in [Ba/H] (cf. Fig.~9f and Fig.~10f in that paper).   
Embarrassingly, these figures suggest that Ba is considerably underabundant
(by $\sim -0.5$~dex or more) in comparison with the solar Ba abundance\footnote{
In this study, Anders \& Grevesse's (1989) solar photospheric Ba abundance 
of $A_{\odot}$(Ba) = 2.13 (in the usual normalization of H = 12.00) is adopted 
in order to keep consistency with Kurucz's (1993) ATLAS9/WIDTH9 
program, which is not much different from the more recent Asplund et al.'s (2009) 
value of $A_{\odot}$(Ba) = $2.18 \pm 0.09$, 
though Gallagher et al. (2020) recently derived a rather higher value of 
$A_{\odot}$(Ba) = 2.27 based on their detailed 3D NLTE analysis because of positive
3D corrections amounting to 0.1--0.2~dex.
Meanwhile, regarding the solar Fe abundance (the reference of metallicity), 
the widely accepted value of $A_{\odot}$(Fe) = 7.50 (Asplund et al. 2009) is used 
as done in the author's previous studies.} for presumably normal stars 
of [Fe/H]~$\sim 0$ as well as for rapid rotators ($v_{\rm e}\sin i \gtrsim 100$~km~s$^{-1}$).
Such A-type stars with appreciably subsolar Ba (even more deficient down to 
[Ba/H]~$\sim -1$) were also identified in the successive Na abundance study
of Takeda et al. (2009), where Ba abundances were also derived as by-products 
(cf. Fig.~8 therein). 
Since an existence of such A stars of near-normal metallicity showing significant 
underabundances of Ba is unusual and rather hard to accept, 
we speculate that something was inadequate 
in the procedure of abundance determination adopted in these papers (i.e., 
spectrum-fitting analysis applied to the region comprising Ba~{\sc ii} 6141 line
under the assumption of Local Thermodynamic Equilibrium: hereinafter LTE).

Above all, the classical assumption of LTE (i.e., neglect of the non-LTE effect) 
may be counted as a likely source of inadequacy. We should bear in mind that only 
Ba~{\sc ii} lines are visible in the spectra of A(--F--G) stars, because the population 
of Ba~{\sc i} (very low ionization potential of $\chi_{\rm ion}^{\rm I}$ = 5.2~eV) is 
utterly negligible (Fig.~1a) while lines of Ba~{\sc iii} (closed shell) are unavailable. 
Here, since the population of Ba~{\sc ii} ($\chi_{\rm ion}^{\rm II} = 10.0$~eV) 
turns into minor species around mid-to-early A stars ($T_{\rm eff} \gtrsim 8000$~K) 
due to enhanced ionization (Fig.~1d), non-LTE overionization effect may become 
significant in such higher $T_{\rm eff}$ regime of strong UV radiation.

\setcounter{figure}{0}
\begin{figure}
\begin{minipage}{80mm}
\includegraphics[width=8.0cm]{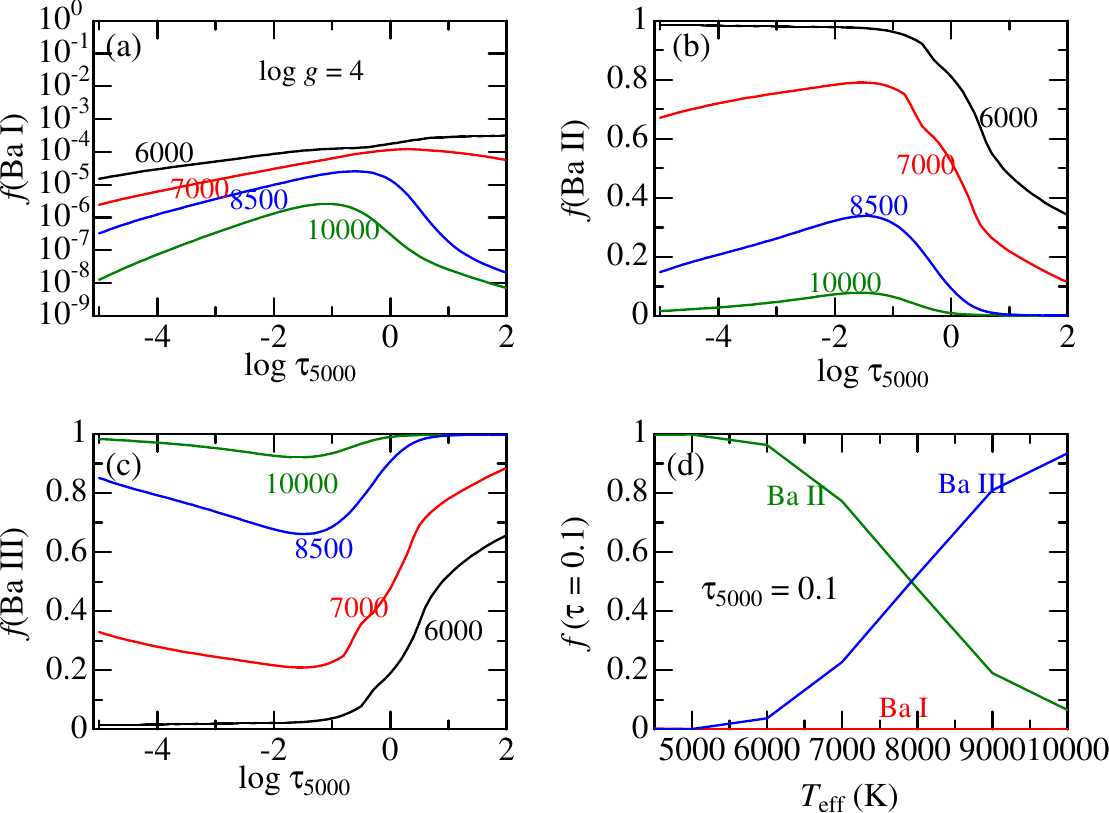}
\caption{
Number population fraction ($f$) of (a) neutral, (b) once-ionized, 
and (c) twice-ionized barium species relative to the total Ba atoms
[e.g., $f$(Ba~{\sc ii}) $\equiv N$(Ba~{\sc ii})/$N_{\rm total}^{\rm Ba}$],
plotted against the continuum optical depth at 5000~\AA. 
Calculations were done for four $\log g = 4.0$ model of different
$T_{\rm eff}$ (6000, 7000, 8500, and 10000~K) as indicated in each panel.
The runs of $f$ for these three stages at $\tau_{5000} = 0.1$ with $T_{\rm eff}$ 
are also depicted in panel (d). All these calculations were done in LTE
(use of Saha's equation).
}
\label{fig1}
\end{minipage}
\end{figure}

Not a few studies of non-LTE Ba~{\sc ii} line formation have been published
so far, but most of them are intended for an application to  
late-type stars of lower $T_{\rm eff}$ (such as solar-type metal-deficient stars, 
extremely metal-poor stars, red giants. etc.); e,g., Mashonkina \& Z\v{a}cs (1996); 
Mashonkina \& Bikmaev (1996); Mashonkina et al. (1999), Short \& Hauschildt (2006);
Andrievsky et al. (2009); Korotin et al. (2011); Dobrovolskas et al. (2012); 
Korotin et al. (2015); Gallagher et al. (2020); 
Liu et al. (2020). 

In contrast, Ba abundance determinations for A-type stars based on non-LTE analysis 
of Ba~{\sc ii} lines are rather scarce in number. To the author's knowledge, available 
publications of NLTE calculations for Ba~{\sc ii} 
are the pioneering work of Gigas (1988) 
for Vega and Mashonkina et al.'s (2020) recent investigation for seven A-type stars 
(HD~32115, HD~73666, Vega, HD~72660, HD~145788, Sirius, and 21~Peg) in the $T_{\rm eff}$ range 
of $\sim$~7200--10400~K.
Yet, these studies are not necessarily useful for application to other 
stars in general, because they are directed to analyzing specific objects and 
how the non-LTE correction generally depends upon atmospheric parameters is not clear.
Therefore, more extensive non-LTE calculations for Ba~{\sc ii} on a large grid 
of model atmospheres may be required to analyze stars spanning a wide range of parameters.

Another potential inadequacy in the author's previous work (Takeda et al. 2008, 2009) 
is that the Ba~{\sc ii} line (at 6141.713~\AA) was treated as a purely single
component. Actually, Ba atoms comprise different stable isotopes  
($^{130}$Ba, $^{132}$Ba, $^{134}$Ba, $^{135}$Ba, $^{136}$Ba, $^{137}$Ba, $^{138}$Ba),
the line wavelengths of which are slightly different from each other (isotope shift).
Furthermore, lines of odd isotopes ($^{135}$Ba, $^{137}$Ba) show complex splitting 
due to the effect of nuclear spin (hyper-fine structure). 
Therefore, Ba abundance derived on the assumption of a single non-split line 
may be overestimated unless a line is weak and unsaturated. 
Since Ba~{\sc ii} lines are generally strong ($W \gtrsim 100$~m\AA) in stars 
under study, this splitting effect should be properly taken into account.
  
The aim of this study is to investigate and clarify the nature of Ba abundance 
anomaly in A-type stars based on a new refined analysis by taking into account the 
non-LTE effect and the hyper-fine structure of line opacity in abundance determination.
Regarding the program stars to be analyzed, attention is paid to Hyades cluster 
stars. where not only A-type stars but also F-, and G-type stars are included.
Since F--G stars are expected to retain the primordial Ba composition at the time 
of star formation, their abundances may serve as the reference for measuring the 
extent of acquired peculiarity in A stars.

In addition, since early A-type stars of comparatively higher $T_{\rm eff}$ are 
insufficient in the Hyades stars, Ba abundances of field A-type stars 
with available observational data are also determined. This supplementary
analysis is separately described in Appendix A.  

\section{Observational data}

The main targets of this investigation are the main-sequence stars of A--F--G 
types belonging to the Hyades cluster, for which high-dispersion spectra are
already available from the previous studies of the author's group.

Regarding A-type stars, 23 Hyades members were selected from the sample of 
122 stars in the Na abundance study of Takeda et al. (2009; cf. 
column 17 of Table~1 therein), which was based on the spectra (covering 
3900--9100~\AA\ with a resolving power of $R \sim 45000$) obtained in 
2008 January, 2008 September, and 2009 January by using the 1.8~m reflector 
along with BOES (Bohyunsan Observatory Echelle Spectrograph) at Bohyunsan 
Optical Astronomy Observatory (BOAO) (see Sect.~2 in Takeda et al. 2009 
for more details).

Takeda et al. (2013) determined the Li, C, and O abundances of 68 Hyades 
F--G stars based on the $R \sim 67000$ spectra (covering 5950--7170~\AA\ region)  
obtained in 2003 December and 2004 March by using HIDES (HIgh Dispersion 
Echelle Spectrograph) at the coude focus of the 188~cm reflector at Okayama 
Astrophysical Observatory (OAO) (see Sect.~2 in Takeda et al. 2013 for more details).
An inspection of these spectra of 68 stars revealed that 2 stars (HD~27483
and HD~28363) show satellite line features indicative of double-line binaries. 
Therefore, excluding these two stars, we adopted the spectra of 66 Hyades F--G 
stars.

As such, 89 Hyades cluster stars (23 A stars and 66 F--G stars) were selected 
as the program stars with the observational data of Takeda et al. (2009, 2013). 
Besides, Procyon (HD~61421) and Sun (Moon)  were chosen as the comparison stars,
for which the OAO/HIDES spectra published by Takeda et al. (2005a) were employed.  

We adopt Ba~{\sc ii} lines at 6141 and 6496~\AA\ for Ba abundance determination 
in this investigation, because these two lines are measurable in the available 
spectra for all of the program stars.\footnote{
The Ba~{\sc ii} 4554 line could not be used because the OAO/HIDES spectra of 68 Hyades 
F--G stars do not cover the blue region.
} Since the wavelength region comprising 
the latter Ba~{\sc ii} 6496 line is contaminated by telluric water vapor lines,
they were removed by dividing the raw spectrum by that of a rapid rotator 
($\alpha$~Leo or $\alpha$~Aql) in application of the IRAF\footnote{
IRAF is distributed by the National Optical Astronomy Observatories, which is 
operated by the Association of Universities for Research in Astronomy, Inc.
under cooperative agreement with the National Science Foundation.
} task ``telluric''  (see, e.g., Sect.~4.1 in Takeda et al. 2009 or Fig.~2 in Takeda et al. 2013).

\section{Atmospheric parameters}

As for the atmospheric parameters [$T_{\rm eff}$, $\log g$ (surface gravity), 
$v_{\rm t}$ (microturbulence), $v_{\rm e}\sin i$ (projected rotational velocity), 
and $A$(Fe) (Fe abundance)] of the 89 Hyades stars, the same values 
as those derived/employed in the above-mentioned papers were adopted unchanged
(see below for brief descriptions). 

23 Hyades A stars: 
$T_{\rm eff}$ and $\log g$ were evaluated from Str\"{o}mgren's $uvby\beta$ 
photometry, $v_{\rm t}$ was calculated by the analytical $T_{\rm eff}$-dependent 
formula (Takeda et al. 2008), and Fe abundance as well as $v_{\rm e}\sin i$ were 
derived from the spectrum-fitting analysis in the 6140--6170~\AA\ region 
(see Sect.~3 and Table~1 in Takeda et al. 2009).

66 Hyades F--G stars: 
$T_{\rm eff}$ and $\log g$ were taken from ``tablea1.dat'' of de Bruijne et al. 
(2001), who established these parameters by comprehensively combining 
colors, theoretical evolutionary tracks, and Hipparcos parallaxes. 
$v_{\rm t}$ was calculated from $T_{\rm eff}$ and $\log g$ 
by applying Eq.(1) and (2) of Takeda et al. (2013).
Fe abundance and $v_{\rm e}\sin i$\footnote{Note that $v_{\rm e}\sin i$ 
adopted here is the quantity denoted as $v_{\rm M}$ in that paper.} 
were derived from the spectrum-fitting analysis in the 6080--6089~\AA\ region
(see Sect.~3 and Table~1 in Takeda et al. 2013).

These values of adopted parameters are summarized in ``hyadesA.dat'' and ``hyadesFG.dat''
of the online material. Fig.~2a--2d illustrates how $\log g$, $v_{\rm t}$, 
$v_{\rm e}\sin i$, and $A$(Fe) change with $T_{\rm eff}$. 

Regarding ($T_{\rm eff}$, $\log g$, $v_{\rm t}$) of Procyon and Sun, 
(6612~K, 4.00, 2.0~km~s$^{-1}$) and (5780~K, 4.44, 1.0~km~s$^{-1}$) 
were adopted by following Takeda et al. (2005b), respectively.   
The model atmosphere for each star was generated by interpolating
Kurucz's (1993) ATLAS9 model grid in terms of $T_{\rm eff}$ and $\log g$,
where solar-metallicity models were employed as in Takeda et al. (2009, 2013).

\setcounter{figure}{1}
\begin{figure}
\begin{minipage}{80mm}
\begin{center}
\includegraphics[width=6.5cm]{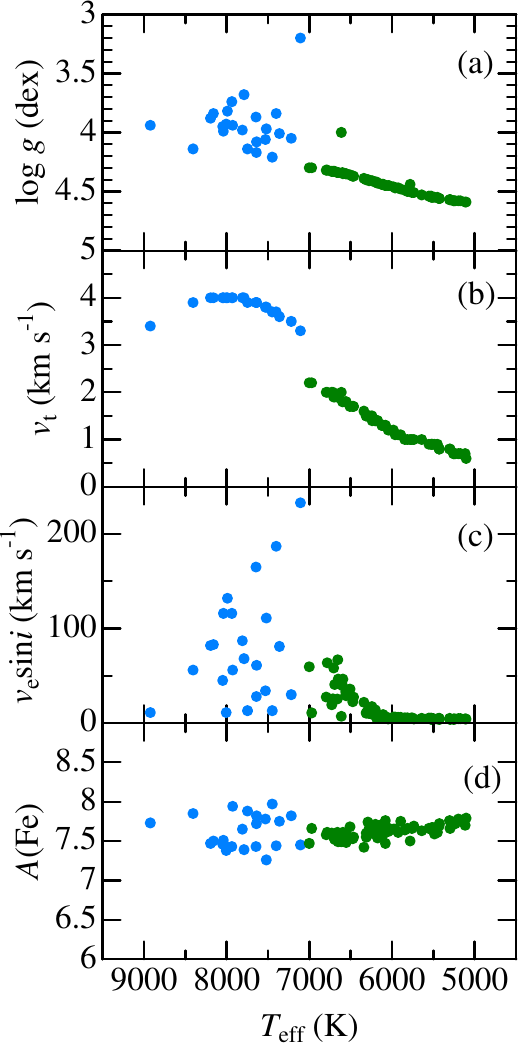}
\caption{
The atmospheric parameters of the program stars adopted
in this study are plotted against $T_{\rm eff}$.
(a) $\log g$ (surface gravity), (b) $v_{\rm t}$ (microturbulence),
(c) $v_{\rm e}\sin i$ (projected rotational velocity), 
and (d) $A$(Fe) (logarithmic number abundance of Fe in the usual 
normalization of H=12). The data of Hyades A stars and F--G stars 
are plotted in blue symbols and green symbols, respectively.  
}
\label{fig2}
\end{center}
\end{minipage}
\end{figure}

\section{Spectrum fitting and equivalent widths}

\setcounter{table}{0}
\begin{table}
\begin{minipage}{90mm}
\normalsize
\caption{Outline of spectrum-fitting analysis.}
\begin{center}
\begin{tabular}{cccc}\hline\hline
Star type & Region (\AA) & Abundances varied$^{*}$ & Figure \\
\hline
\multicolumn{4}{c}{(Ba~{\sc ii} 6141 region)}\\
A stars    & 6130--6152 & Si, Fe, Ba     & Fig.~3a \\
F--G stars & 6135--6150 & Si, Fe, Ba     & Fig.~3b--d \\
\hline
\multicolumn{4}{c}{(Ba~{\sc ii} 6496 region)}\\
A stars    & 6480--6510 & Ca, Ti, Fe, Ba & Fig.~4a \\
F--G stars & 6485--6505 & Ca, Ti, Fe, Ba & Fig.~4b--d \\
\hline
\end{tabular}
\end{center}
\scriptsize
$^{*}$ The abundances of all other elements than these were fixed in the fitting. 
\end{minipage}
\end{table}

\setcounter{table}{1}
\begin{table*}
\begin{minipage}{180mm}
\begin{center}
\caption{Atomic data of Ba~{\sc ii} lines with hyper-fine structures adopted in this study.}
\normalsize
\begin{tabular}{c@{  }c@{  }c c@{  }c@{  }c c@{  }c@{  }c }
\hline\hline                 
\multicolumn{3}{c}{Ba~{\sc ii} 4554 (17 components)} & \multicolumn{3}{c}{Ba~{\sc ii} 6141 (23 components)} & 
\multicolumn{3}{c}{Ba~{\sc ii} 6496 (17 components)} \\
\multicolumn{3}{c}{$^{2}{\rm S}_{1/2}$--$^{2}{\rm P}_{3/2}^{\circ}$ (multiplet 1)} & 
\multicolumn{3}{c}{$^{2}{\rm D}_{5/2}$--$^{2}{\rm P}_{3/2}^{\circ}$ (multiplet 2)} & 
\multicolumn{3}{c}{$^{2}{\rm D}_{3/2}$--$^{2}{\rm P}_{1/2}^{\circ}$ (multiplet 2)} \\
\multicolumn{3}{c}{$\chi_{\rm low} = 0.000$~eV} & \multicolumn{3}{c}{$\chi_{\rm low} = 0.704$~eV} & 
\multicolumn{3}{c}{$\chi_{\rm low} = 0.604$~eV} \\
\multicolumn{3}{c}{$\log (\Sigma gf) = +0.170$} & \multicolumn{3}{c}{$\log (\Sigma gf) = -0.076$} & 
\multicolumn{3}{c}{$\log (\Sigma gf) = -0.377$} \\
\multicolumn{3}{c}{Gammar = 8.20} & \multicolumn{3}{c}{Gammar = 8.20} & 
\multicolumn{3}{c}{Gammar = 8.10} \\
\multicolumn{3}{c}{Gammas = $(-5.82)$} & \multicolumn{3}{c}{Gammas = $(-5.82)$} & 
\multicolumn{3}{c}{Gammas = $(-5.85)$} \\
\multicolumn{3}{c}{Gammaw = $-7.65$} & \multicolumn{3}{c}{Gammaw = $-7.58$} & 
\multicolumn{3}{c}{Gammaw = $-7.58$} \\
\hline
$\lambda$ & $\log gf$ & isotope &  $\lambda$ & $\log gf$ & isotope & $\lambda$ & $\log gf$ & isotope \\
\hline
 4553.9980 & $-$1.586& 137& 6141.7086 & $-$2.260& 137& 6496.8826 & $-$2.832& 137  \\
 4553.9991 & $-$1.586& 137& 6141.7086 & $-$1.452& 137& 6496.8857 & $-$3.063& 135  \\
 4553.9993 & $-$1.984& 137& 6141.7094 & $-$3.406& 137& 6496.8873 & $-$2.133& 137  \\
 4554.0009 & $-$1.817& 135& 6141.7105 & $-$1.683& 135& 6496.8901 & $-$2.364& 135  \\
 4554.0021 & $-$1.817& 135& 6141.7107 & $-$2.491& 135& 6496.8959 & $-$1.686& 137  \\
 4554.0024 & $-$2.215& 135& 6141.7113 & $-$3.637& 135& 6496.8977 & $-$1.917& 135  \\
 4554.0310 & $-$2.805& 130& 6141.7129 & $-$0.220& 138& 6496.8977 & $-$0.521& 138  \\
 4554.0315 & $-$2.826& 132& 6141.7140 & $-$1.181& 136& 6496.8988 & $-$1.482& 136  \\
 4554.0319 & $-$1.447& 134& 6141.7146 & $-$1.658& 137& 6496.9004 & $-$1.994& 134  \\
 4554.0323 & $-$0.935& 136& 6141.7154 & $-$2.163& 137& 6496.9015 & $-$2.531& 137  \\
 4554.0336 & +0.026& 138& 6141.7154 & $-$1.693& 134& 6496.9021 & $-$3.373& 132  \\
 4554.0479 & $-$1.370& 135& 6141.7157 & $-$1.889& 135& 6496.9024 & $-$2.762& 135  \\
 4554.0503 & $-$1.139& 137& 6141.7163 & $-$3.230& 137& 6496.9035 & $-$2.133& 137  \\
 4554.0506 & $-$1.817& 135& 6141.7163 & $-$2.394& 135& 6496.9037 & $-$3.352& 130  \\
 4554.0518 & $-$2.516& 135& 6141.7169 & $-$3.072& 132& 6496.9044 & $-$2.364& 135  \\
 4554.0536 & $-$1.586& 137& 6141.7170 & $-$3.461& 135& 6496.9082 & $-$2.133& 137  \\
 4554.0547 & $-$2.285& 137& 6141.7174 & $-$1.908& 137& 6496.9088 & $-$2.364& 135  \\
           &       &    & 6141.7183 & $-$3.051& 130&           &       &      \\
           &       &    & 6141.7184 & $-$2.276& 137&           &       &      \\
           &       &    & 6141.7184 & $-$2.139& 135&           &       &      \\
           &       &    & 6141.7188 & $-$2.230& 137&           &       &      \\
           &       &    & 6141.7191 & $-$2.507& 135&           &       &      \\
           &       &    & 6141.7198 & $-$2.461& 135&           &       &      \\
\hline
\end{tabular}
\end{center}
The meanings of most of these data are self-explanatory. 
See Table~2 of Takeda et al. (2009) for the definition of the damping parameter: 
radiation damping (Gammar), Stark effect damping (Gammas), and van der Waals effect 
damping (Gammaw). 
All data were taken from VALD (Ryabchikova et al. 2015), except for those of Gammar 
(evaluated from the Einstein coefficients of spontaneous transition for the upper 
and lower levels) and those parenthesized of Gammas (calculated by the default 
treatment of WIDTH9 program). These $gf$(VALD) values are scaled by assuming the
isotopic composition of the solar system. 
Note that only Ba~{\sc ii} 6141 and 6496 lines (multiplet~2) are employed for  
Ba abundance determinations in this study, while the Ba~{\sc ii} 4554 line
(multiplet~1) is used only for the discussing the characteristic tendency of 
the non-LTE effect in Sect.~5.3 (i.e., Fig.~5 and Fig.~6).
\end{minipage}
\end{table*}

The procedures of spectrum fitting follow by equivalent width derivation (done with 
the assumption of LTE at this stage) are essentially the same as adopted in the author's 
previous studies (see, e.g., Sect.~5 in Takeda et al. 2013).
 
First, multi-parameter fitting technique is applied to two spectral regions comprising 
Ba~{\sc ii} 6141 and 6496 lines as outlined in Table~1. The adopted atomic data of 
Ba~{\sc ii} lines (including isotope shifts and hyperfine structures) are given in Table~2, 
while line data of all other elements are also taken from VALD (Ryabchikova et al. 2015). 
The finally accomplished fit between the theoretical spectrum (corresponding to 
the converged solutions of parameters) and the observed spectrum for each star 
is depicted in Fig.~3 (6141 region) and Fig.~4 (6496 region). 

Then, the equivalent width of the Ba~{\sc ii} 6141 and 6496 lines ($W_{6141}$ and 
$W_{6496}$) was inversely evaluated from the abundance solution (resulting 
from the fitting analysis) by using the WIDTH9 program (Kurucz 1993) which was
considerably modified by the author (e.g., treatment of multi-component lines,
incorporating non-LTE departure coefficients, etc). These $W$ values are further 
employed to derive non-LTE abundances and to estimate abundance sensitivities to 
parameter changes (Sect.~6)

\setcounter{figure}{2}
\begin{figure*}
\begin{minipage}{160mm}
\begin{center}
\includegraphics[width=12.0cm]{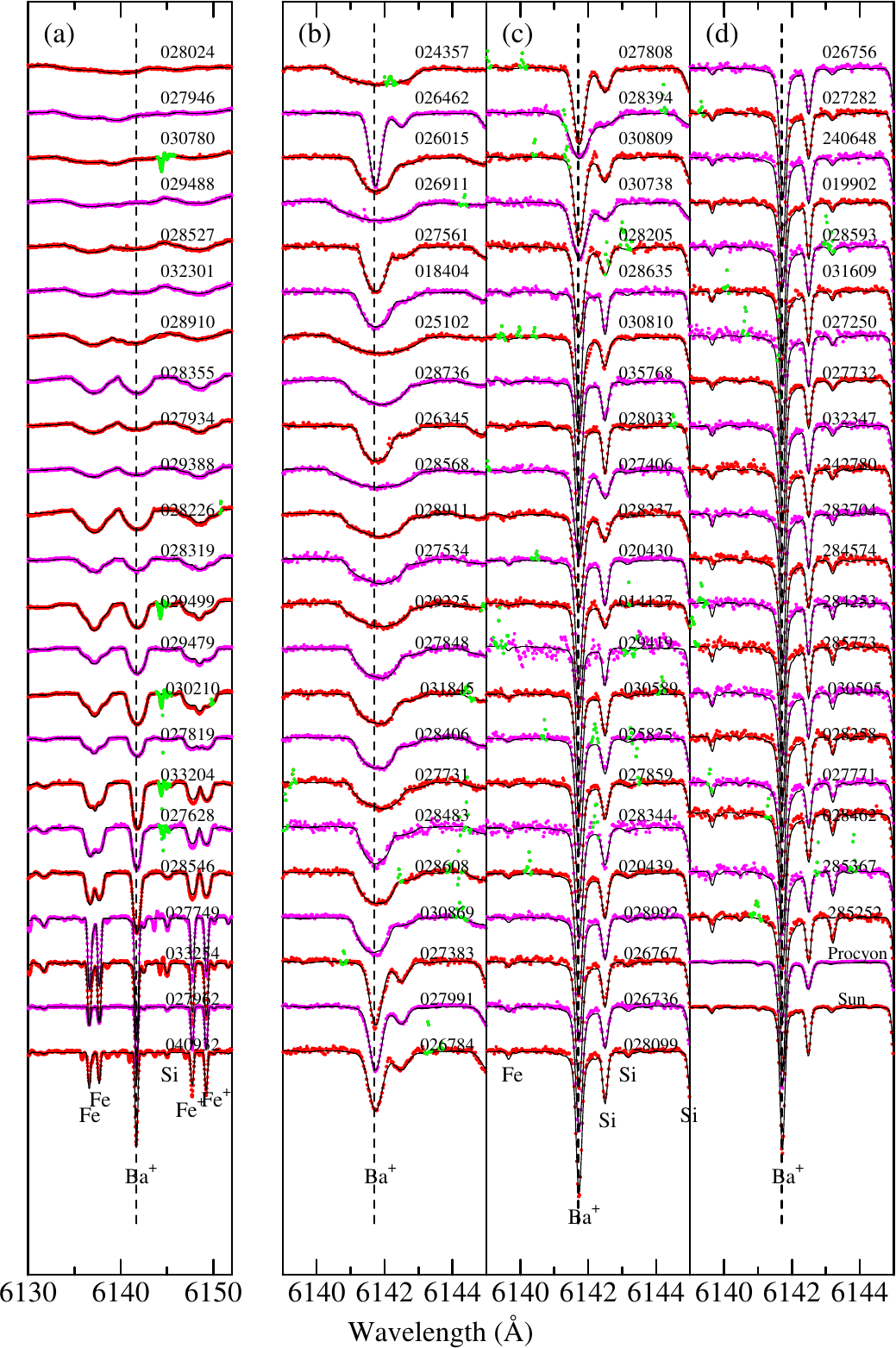}
\caption{
Synthetic spectrum fitting around the Ba~{\sc ii} 6141 line region for Ba 
abundance determination. The best-fit theoretical spectra are depicted by 
black solid lines, while the observed data are plotted by colored symbols
(inappropriate rejected data in evaluating the goodness of fit are 
highlighted in light green).
The leftmost panel (a) is for 23 Hyades A-type stars arranged in the descending
order of $v_{\rm e}\sin i$. Panels (b), (c), and (d) in the right-hand side 
are for 66 Hyades F- and G-type stars (along with Procyon and Sun) arranged 
in the descending order of $T_{\rm eff}$ (from top to bottom; from left to right).
HD number is indicated in each spectrum. 
An offset of 0.2 (in unit of the continuum-normalized flux) is applied to 
each spectrum relative to the adjacent one. The position of the Ba~{\sc ii} 6141 
line is indicated by vertical dashed lines.
Note that, while the whole wavelength region adopted for fitting (cf. Table~1) 
is shown for A stars in panel (a), only the restricted region around the 
Ba~{\sc ii} line is displayed for F--G stars in panels (b)--(d).
}
\label{fig3}
\end{center}
\end{minipage}
\end{figure*}

\setcounter{figure}{3}
\begin{figure*}
\begin{minipage}{160mm}
\begin{center}
\includegraphics[width=12.0cm]{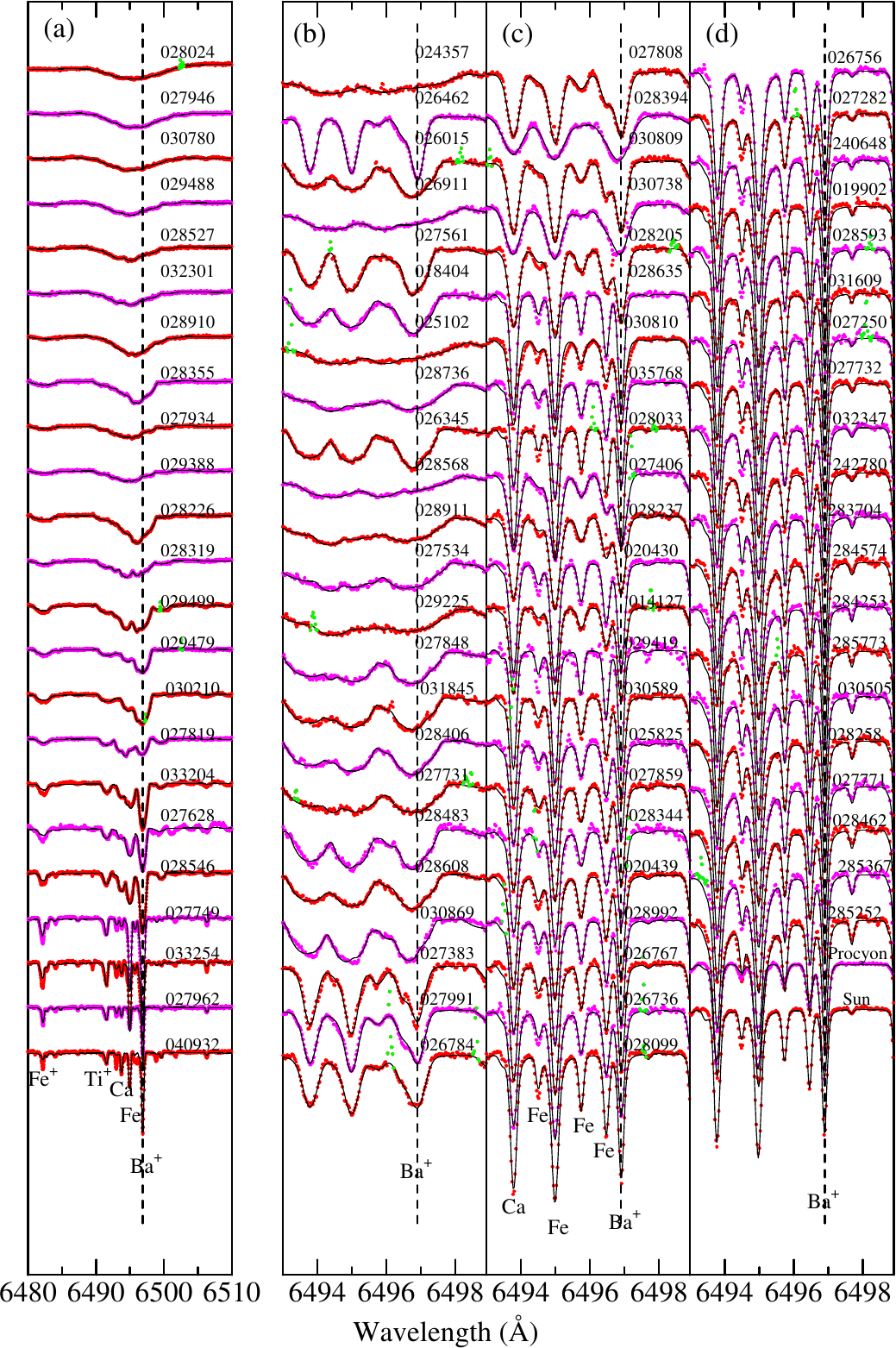}
\caption{
Synthetic spectrum fitting around the Ba~{\sc ii} 6496 line region for Ba 
abundance determination. The position of the Ba~{\sc ii} 6496 line is 
indicated by vertical dashed lines. Otherwise, the same as in Fig.~3.
}
\label{fig4}
\end{center}
\end{minipage}
\end{figure*}

\section{Statistical equilibrium calculation for Ba~{\sc ii}}

\subsection{Atomic model of barium}

The non-LTE calculations for Ba~{\sc ii} were carried out based on the 
Ba~{\sc ii} model atom comprising 39 terms (up to 5p$^{6}$\,14f\;$^{2}$F 
at 78163 cm$^{-1}$) and 258 radiative transitions, which was constructed 
by consulting the atomic line data 
(filename: ``gf5601.all'')\footnote{http://kurucz.harvard.edu/atoms/5601/}
 compiled by Dr. R. L. Kurucz. Though the contribution of Ba~{\sc i} was neglected,
the ground level of Ba~{\sc iii} was taken into account in the number conservation of 
total Ba atoms. 

Regarding the photoionization cross section, the data read from Fig.~1 of 
Mashonkina et al. (2020) were used for the lowest 3 terms (6s\,$^{2}$S, 5d\,$^{2}$D, 
and 6p\,$^{2}$P$^{\circ}$), while the hydrogenic approximation was assumed for 
the remaining terms.  Otherwise (such as the treatment of collisional rates), 
the recipe described in Sect.~3.1.3 of Takeda (1991) was followed (inelastic 
collisions due to neutral hydrogen atoms were formally included as described 
therein without any correction).   

\subsection{Grid of models}

The calculations were done on a grid of 48 ($= 12 \times 4$) 
solar-metallicity model atmospheres 
resulting from combinations of twelve $T_{\rm eff}$ values 
(5000, 5500, 6000, 6500, 7000, 7500, 8000, 8500, 9000, 9500, 10000, 10500~K) 
and four  $\log g$ values (3.0, 3.5, 4.0, and 4.5).
while assuming $v_{\rm t}$ = 2~km~s$^{-1}$.
Regarding the input Ba abundance, three values of $A$(Ba) = 2.13,
3.13, and 4.13 (corresponding to [Ba/H] = 0, +1, and +2) were assumed  
(resulting in three kinds of non-LTE grids), in order to address
stars of considerable Ba excess (cf. Sect.~6.1). 
The depth-dependent non-LTE departure coefficients to be used for 
each star were then evaluated by interpolating the grid in terms 
of $T_{\rm eff}$ and $\log g$.

\subsection{Characteristic trend of the non-LTE effect}

Fig.~5 displays $l_{0}^{\rm NLTE}(\tau)/l_{0}^{\rm LTE}(\tau)$ 
(non-LTE-to-LTE line-center opacity ratio) and 
$S_{\rm L}(\tau)/B(\tau)$ (ratio of the line source function to the Planck 
function) for the transition relevant to the Ba~{\sc ii} 4554 line of multiplet 1 
(upper set) and Ba~{\sc ii} 6141/6496 lines of multiplet 2 (lower set) 
as functions of optical depth at 5000\AA\ for selected representative cases. 
Likewise, Fig.~6 illustrates how the theoretical equivalent widths calculated in LTE 
($W^{\rm L}$) as well as in non-LTE ($W^{\rm N}$) and the corresponding non-LTE 
corrections ($\Delta \equiv A^{\rm N} - A^{\rm L}$, where $A^{\rm L}$ and $A^{\rm N}$ are 
the abundances derived from $W^{\rm N}$ with LTE and non-LTE) depend upon $T_{\rm eff}$.
The following characteristic trends are read from these figures (since the results for 
4554 line and 6141/6496 lines are not much different, we focus here only on the latter 
multiplet 2 lines actually used for Ba abundance determination in this study).
\begin{itemize}
\item
As clearly seen from Fig.~6, the trend of the non-LTE effect in  A stars 
($T_{\rm eff} \gtrsim 7500$~K) and that in F--G stars ($T_{\rm eff} \lesssim 7500$~K)
are distinctly different.
\item 
Regarding A-type stars, the inequality $W^{\rm N} < W^{\rm L}$ (and $\Delta > 0$) 
generally holds (i.e., non-LTE line weakening) and this tendency  becomes progressively
conspicuous as $T_{\rm eff}$ increases from $\sim 7500$~K ($\Delta \sim 0$) 
to $\sim 9000$~K (where $\Delta$ attains a maximum as large as $\sim +0.5$~dex).
This is due to the decrease of line opacity caused by overionization of the 
lower level population (cf. Fig.~5), which becomes manifest only at higher 
$T_{\rm eff}$ ($> 7000$~K) where UV radiation is enhanced.
\item
In contrast, the situation is inversed in F--G stars ($T_{\rm eff} \lesssim 7500$~K)  
where the inequality makes $W^{\rm N} > W^{\rm L}$ (and $\Delta < 0$) indicating
non-LTE line-strengthening. This effect gradually increases as $T_{\rm eff}$ is
lowered from $\sim 7500$~K to $\sim 6500$~K (where $|\Delta|$ attains a maximum 
amounting to $\sim$~0.2~dex), and then turns to decrease towards $\sim 5000$~K (Fig.~6). 
In this lower $T_{\rm eff}$ regime, the Ba~{\sc ii} line is intensified by the non-LTE
effect because of the dilution of line source function due to escape of line photons
($S_{\rm L}/B < 1$; cf. Fig.~5) which makes the line core deeper. 
\item
As to $g$-dependence of the non-LTE effect, $|\Delta|$ tends to be larger for lower 
$\log g$ (Fig.~6). This is reasonably understood because a decrease of surface 
gravity leads to a lower density atmosphere of more enhanced departure from LTE.
\item
Quantitatively, the non-LTE effect is comparatively important in A-type stars with 
positive $\Delta$ values (tending to increase with $T_{\rm eff}$) amounting 
up to $\sim 0.5$~dex at early A, while negative non-LTE corrections 
for the case of F--G stars ($|\Delta| \sim 0.2$~dex at most) are less significant.
\item
These trends are almost consistent (at least qualitatively) with the consequences
of previous non-LTE studies of Ba~{\sc ii} line formation (cf. Sect.~1).
For example, our non-LTE corrections for the Ba~{\sc ii} 6141 line 
derived for $\log g= 4$ models of $T_{\rm eff}$ = 5000, 5500, 6000, and 6500~K 
are $-0.07$, $-0.11$, $-0.16$, and $-0.18$~dex (for the [Fe/H] = [Ba/H] = 0 case), 
are favorably compared with Korotin et al.'s (2015) corresponding values 
of $-0.04$, $-0.07$, $-0.10$, and $-0.11$~dex (read  from their  ``table5.dat'').
Regarding A-type stars, comparing our $\Delta_{6141}$ values [$\sim -0.2$~dex 
($T_{\rm eff} \sim$~7000~K), +0.1--0.2~dex ($T_{\rm eff} \sim$~8000~K) and +0.4--0.5~dex 
($T_{\rm eff} \sim$~9000--10000~K)] with Table~2 of Mashonkina et al. (2020), we can see 
that the qualitative tendency is the same but their results appear to be quantitatively 
somewhat smaller especially in the regime of higher $T_{\rm eff}$ (i.e., +0.1--0.4~dex 
at $T_{\rm eff} \sim$~9000--10000~K).
\end{itemize}

\setcounter{figure}{4}
\begin{figure}
\begin{minipage}{90mm}
\includegraphics[width=9.0cm]{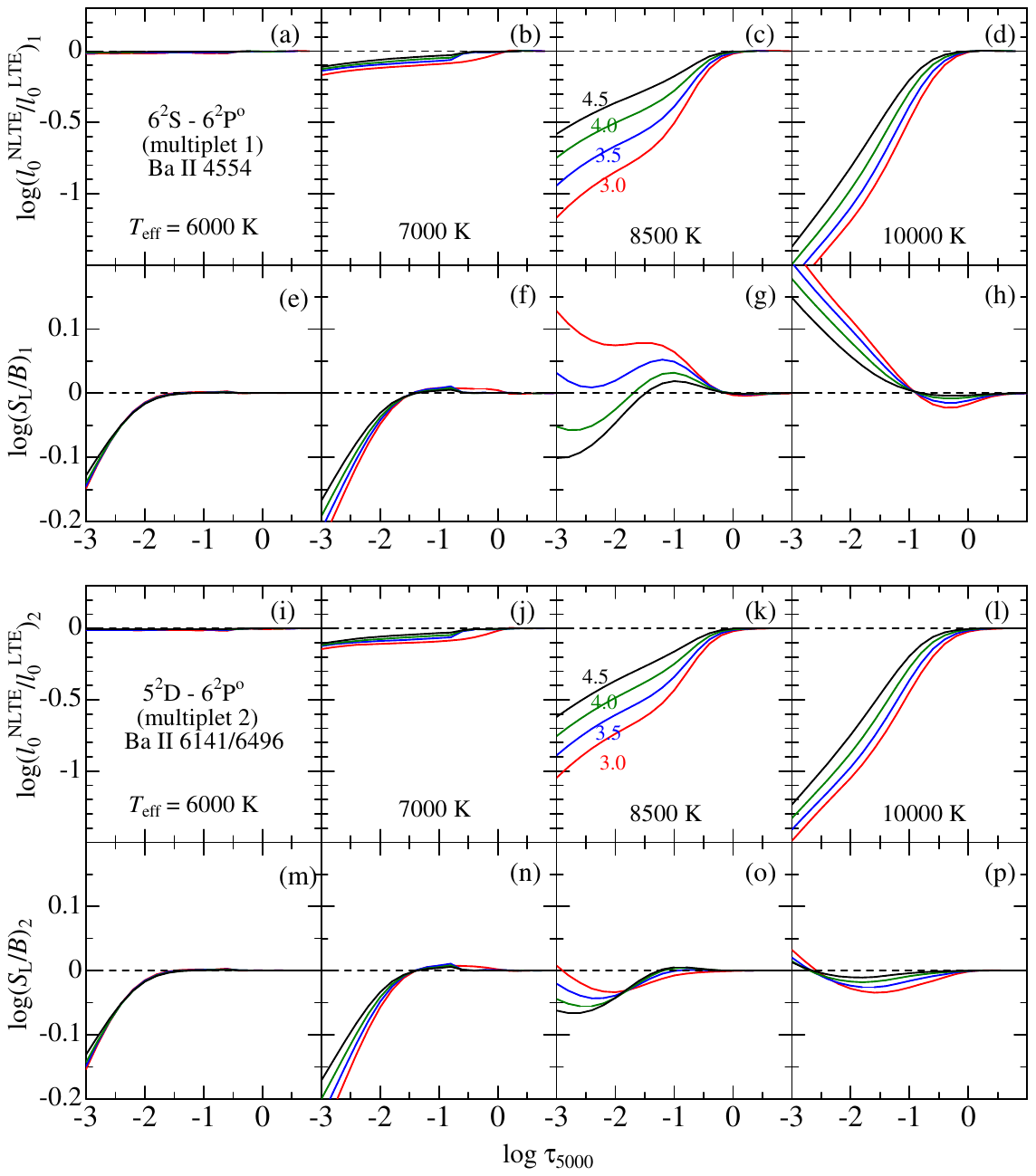}
\caption{
Non-LTE-to-LTE line-center opacity ratio (upper row) and 
the ratio of the line source function ($S_{\rm L}$) 
to the local Planck function ($B$) (lower row)  
plotted against the continuum optical depth at 5000~\AA. 
Shown here are the results of calculations done with $v_{\rm t} = 2$~km~s$^{-1}$ 
and solar Ba abundance ([Ba/H] = 0) on the solar-metallicity models 
of $T_{\rm eff} =$ 6000, 7000, 8500, and 10000~K (from left to right).
At each panel, the results for four $\log g$ values of 3.0, 3.5, 4.0, and 4.5 
are depicted by different colors (red, blue, green, and black, respectively). 
The upper set of 8 panels (a)--(h) are for the 
Ba~{\sc ii} 6$^{2}{\rm S}$--6$^{2}{\rm P}^{\circ}$ transition 
of multiplet~1 (corresponding to Ba~{\sc ii} 4554 line), 
while the lower set of 8 panels (i)--(p) are for the
Ba~{\sc ii} 5$^{2}{\rm D}$--6$^{2}{\rm P}^{\circ}$ transition 
of multiplet 2 (corresponding to Ba~{\sc ii} 6141/6496 lines). 
}
\label{fig5}
\end{minipage}
\end{figure}

\setcounter{figure}{5}
\begin{figure}
\begin{minipage}{80mm}
\includegraphics[width=8.0cm]{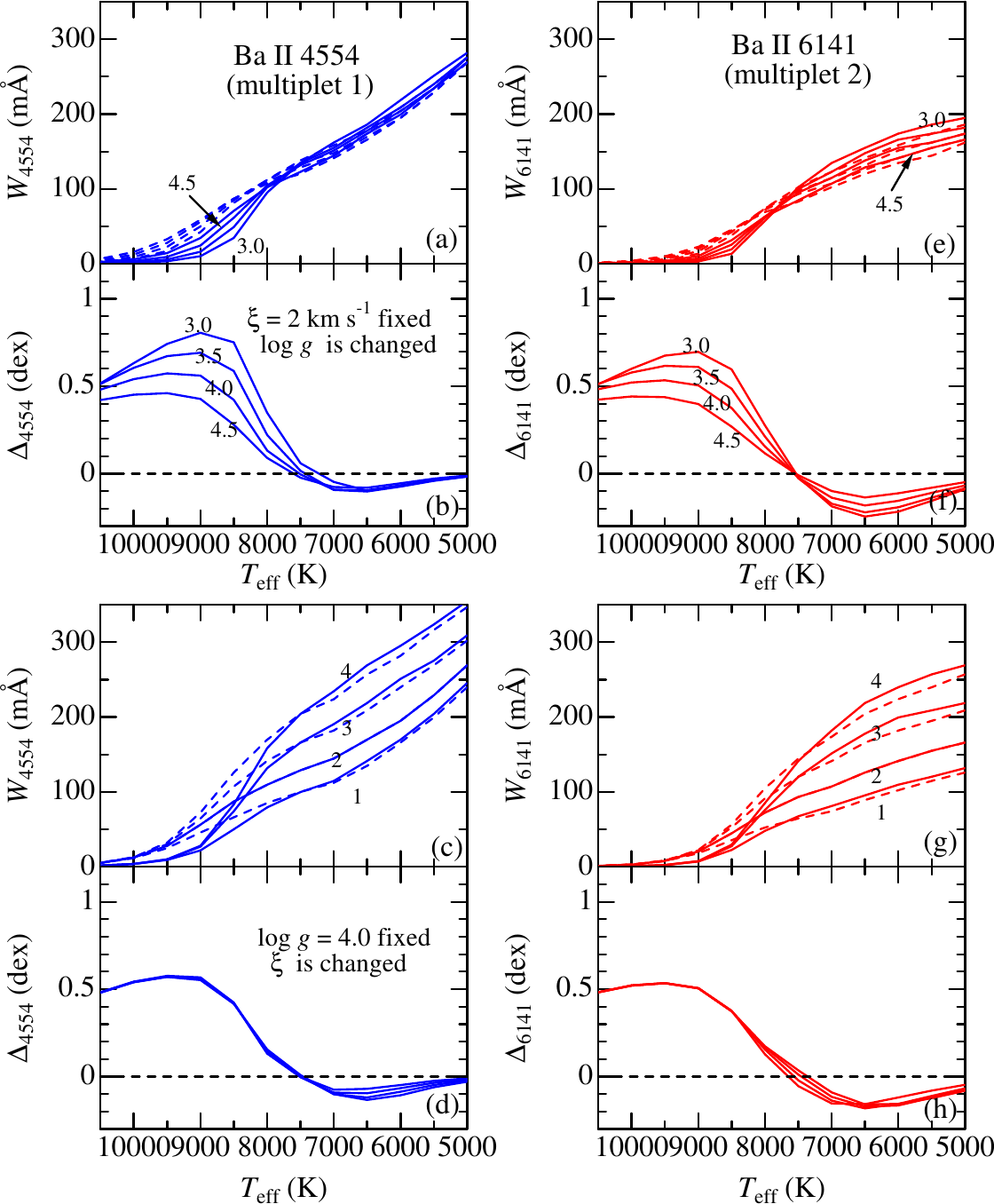}
\caption{
Theoretically calculated non-LTE and LTE equivalent widths 
($W^{\rm N}$ and $W^{\rm L}$) and the corresponding non-LTE corrections 
($\Delta \equiv A^{\rm N} - A^{\rm L}$, where $A^{\rm L}$ and $A^{\rm N}$ 
are the abundances derived from $W^{\rm N}$ with LTE and non-LTE) 
are plotted against $T_{\rm eff}$.
These data were computed with the solar Ba abundance ([Ba/H] = 0) 
on the grid of models described in Sect.~5.2,
The upper set of panels show the results for different $\log g$ 
(3.0, 3.5, 4.0, and 4.5) but fixed $v_{\rm t}$ (2~km~s$^{-1}$),
and the lower set is for the cases of different $v_{\rm t}$
(1, 2, 3, and 4~km~s$^{-1}$) but fixed $\log g$ (4.0). 
In each two-panel set, the upper panel gives $W^{\rm N}$ (solid line) 
and $W^{\rm L}$ (dashed line), while the lower one is for $\Delta$.
The left-column panels (a)--(d) and right-column panels (e)--(h) are for the 
Ba~{\sc ii} 4554 line (multiplet 1) and Ba~{\sc ii} 6141 line (multiplet 2), 
respectively.  
}
\label{fig6}
\end{minipage}
\end{figure}

\section{Determination of non-LTE Ba abundances}

\subsection{Applied non-LTE corrections}

Now, based on the non-LTE calculations done in Sect.~5, $A^{\rm N}$, 
$A^{\rm L}$, and $\Delta$ (NLTE as well as LTE abundances and NLTE correction; 
cf. Sect.~5.3) can be derived from both $W_{6141}$ and $W_{6496}$ (Sect.~4) 
for each of the program stars. Here, since the non-LTE departure coefficients 
are more or less dependent upon the value of [Ba/H] assumed in the calculation 
while stellar Ba abundances tend to be generally diversified, it is necessary 
to adopt a NLTE correction corresponding to an adequate [Ba/H] consistent 
with the final non-LTE abundance. Therefore, we proceed as follows.

Regarding A-type stars showing large star-to-star diversities 
($0 \lesssim$~[Ba/H]~$\lesssim 2$), three kinds of non-LTE abundances are 
derived from $W$ ($A_{0}^{\rm N}$, $A_{+1}^{\rm N}$, and $A_{+2}^{\rm N}$ 
corresponding to [Ba/H] = 0, +1, and +2, respectively) by using three sets 
of departure coefficients prepared for each star (cf. Sect.~5.2).\footnote{
Actually, the sensitivity of $A^{\rm N}$ to [Ba/H] is not very large but comparatively 
mild. For example, the difference between the [Ba/H] = +2 and [Ba/H] = 0 cases 
($|A_{+2}^{\rm N} - A_{0}^{\rm N}|$) is $\lesssim$~0.1--0.2~dex at most.} 
These $A_{0}^{\rm N}$, $A_{+1}^{\rm N}$, and $A_{+2}^{\rm N}$ are sufficient 
to express $A^{\rm N}$ by a second-order polynomial (Lagrange interpolation) 
in terms of $x (\equiv {\rm [Ba/H]})$ as 
\begin{equation}
\begin{split}
A^{\rm N} = & A_{0}^{\rm N} \frac{(x- x_{1})(x - x_{2})}{(x_{0} - x_{1})(x_{0} - x_{2})} + 
   A_{+1}^{\rm N} \frac{(x- x_{0})(x - x_{2})}{(x_{1} - x_{0})(x_{1} - x_{2})} + \\
   & A_{+2}^{\rm N} \frac{(x- x_{0})(x - x_{1})}{(x_{2} - x_{0})(x_{2} - x_{1})},  
\end{split}
\end{equation}
where $x_{0}$, $x_{1}$, and $x_{2}$ are 0, +1, and +2, respectively. 
We may thus write $A^{\rm N}$ for simplicity as
\begin{equation}
A^{\rm N} = a  + b x + c x^{2},
\end{equation} 
where $a$, $b$, and $c$ are known coefficients.
Meanwhile, according to the definition,
\begin{equation}
A^{\rm N} = x + 2.13.  
\end{equation}
Combining Eqs.~(2) and (3), we have 
\begin{equation}
a + b x + c x^{2} = x+ 2.13.
\end{equation} 
By using the solution of Eq.~(4) denoted as $x_{*}$ 
(which of the two solutions should be adopted is self-evident), 
we obtain $A_{*}^{\rm N} (= x_{*} + 2.13)$ and 
$\Delta_{*} (= A_{*}^{\rm N} - A^{\rm L})$   
as the final non-LTE abundance and non-LTE correction.

For the case of F--G stars, Ba abundances are not much 
different from the solar value. Therefore, $A^{\rm N}$ is linearly interpolated 
in terms of $x$ by using two non-LTE abundances ($A_{0}^{\rm N}$ and 
$A_{+1}^{\rm N}$) as
\begin{equation}
A^{\rm N} = 
  A_{0}^{\rm N} \frac{(x - x_{1})}{(x_{0} - x_{1})} + 
  A_{+1}^{\rm N} \frac{(x - x_{0})}{(x_{1} - x_{0})}. 
\end{equation}
Again, we write for simplicity as
\begin{equation}
A^{\rm N} = a  + b x.
\end{equation}
Combining Eqs.~(6) and (3), we have 
\begin{equation}
a + b x = x+ 2.13,
\end{equation} 
and the solution $x_{*}$ gives $A_{*}^{\rm N}$ and $\Delta_{*}$. 

The resulting values of $W$, $A^{\rm L}$, $\Delta$, and $A^{\rm N}$ (the 
subscript ``*'' in $A^{\rm N}$ and $\Delta$ is omitted for simplicity) 
obtained from two Ba~{\sc ii} lines for 23 Hyades A and 66 Hyades F--G 
stars are summarized in ``hyadesA.dat'' and ``hyadesFG.dat'' of the 
supplementary material. How these quantities depend upon $T_{\rm eff}$ 
is also illustrated in Fig.~7 (panels (a)--(d) and panels (h)--(k)), where 
the left and right panels correspond to Ba~{\sc ii} 6141 and 
Ba~{\sc ii} 6496 lines, respectively. 
As seen from Fig.~7o, $A_{6141}^{\rm N}$ and $A_{6496}^{\rm N}$ are well 
correlated with each other, though the former tends to be slightly lower 
by $\sim 0.05$~dex on the average. 

Regarding the reference stars, where $W_{6141}/W_{6496}$ are 129/125 (Procyon) 
and 111/102 (Sun), the results of $A_{6141}^{\rm N}/A_{6496}^{\rm N}$ 
($\Delta_{6141}/\Delta_{6496}$) are 2.08/2.15 ($-0.18/-0.17$) for Procyon,  
and 2.12/2.14 ($-0.08/-0.09$) for the Sun, which are overplotted in Fig.~7. 
Therefore, these $A^{\rm N}$ values derived from Ba~{\sc ii} 6141/6496 lines 
for these two standard stars are in fairly good agreement with the fiducial solar
Ba abundance ($A_{\odot} = 2.13$) adopted in this study (cf. footnote~3).

\setcounter{figure}{6}
\begin{figure}
\begin{minipage}{90mm}
\includegraphics[width=9.0cm]{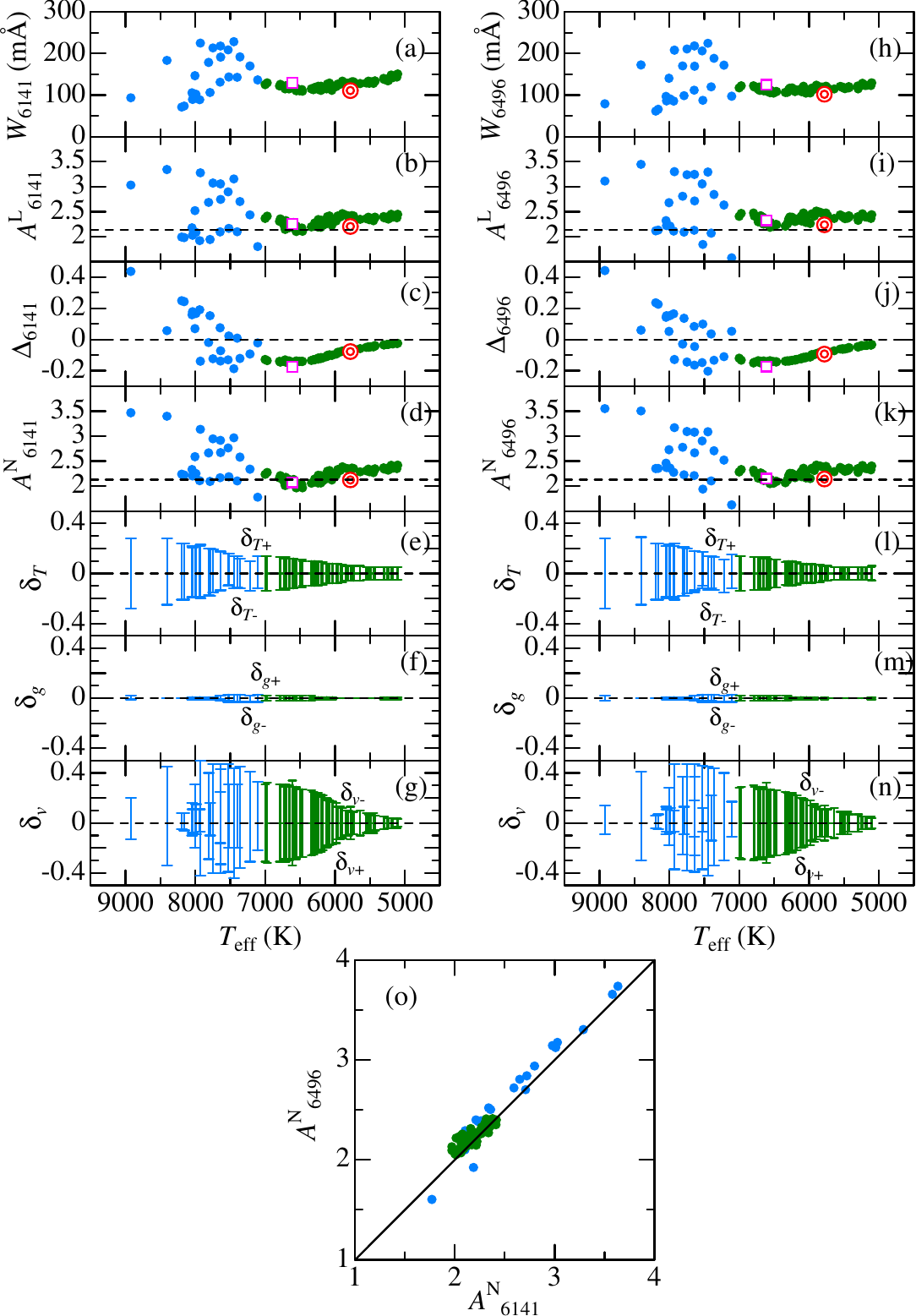}
\caption{
Equivalent widths ($W$), LTE abundances ($A^{\rm L}$), NLTE corrections ($\Delta$),
NLTE-corrected abundances ($A^{\rm N}$) and abundance sensitivities to perturbations 
of atmospheric parameters ($\delta_{T\pm}$, $\delta_{g\pm}$, and $\delta_{v\pm}$; 
cf. Sect.~6.2)  are plotted against $T_{\rm eff}$. The left panels (a)--(g) are for 
Ba~{\sc ii} 6141, and the right ones (h)--(n) are for Ba~{\sc ii} 4696.
How $A_{6141}^{\rm N}$ and $A_{6496}^{\rm N}$ are correlated with each other 
is shown in the bottom panel (o). Hyades A stars and Hyades F--G stars are 
indicated in blue and green colors, respectively. The results for the Sun (red
double circle) and Procyon (pink open square) are also overplotted for comparison.  
Dashed lines in panels (b), (d), (i), and (k) indicate the position of solar 
Ba abundance (2.13).  
}
\label{fig7}
\end{minipage}
\end{figure}

\subsection{Sensitivities to parameter changes}

How the Ba abundances are affected by changing atmospheric parameters ($T_{\rm eff}$, 
$\log g$, and $v_{\rm t}$) was estimated by repeating the analysis on 
the $W$ values while perturbing these standard parameters interchangeably 
by $\pm 3\%$, $\pm 0.1$~dex, and $\pm 30\%$, which were tentatively chosen 
by comprehensively consulting the uncertainties of parameters assumed
in Takeda et al. (2008, 2009) (A stars)  and Takeda et al. (2013) (F--G stars).  
The resulting responses of 
$\delta_{T\pm}$ (abundance changes by perturbations of $T_{\rm eff}$), 
$\delta_{g\pm}$ (abundance changes by perturbations of $\log g$), and
$\delta_{v\pm}$ (abundance changes by perturbations of $v_{\rm t}$)
are illustrated against $T_{\rm eff}$ in Fig.~7e--7g (6141 line)
and Fig.~7l--7n (6496 line). 
An inspection of these figures reveals the following trends.
\begin{itemize}
\item
The most important parameter influencing Ba abundances of
A--F stars is $v_{\rm t}$, which is because the Ba~{\sc ii} 6141/6496 lines 
are so strong as to be saturated ($W \gtrsim 100$~m\AA) in most cases. 
Since empirical analytical formulae were invoked for $v_{\rm t}$ in this study,\footnote{
A large fraction of our program stars (especially A--F stars at 
10000~K~$\gtrsim T_{\rm eff} \gtrsim 6500$~K) show appreciably broad lines 
($v_{\rm e}\sin i$ larger than several tens of km~s$^{-1}$; cf. Fig~2c),
for which the conventional $v_{\rm t}$-determination method (e.g., requiring 
the abundance consistency of many Fe lines of different strengths) is
hardly applicable because directly measuring the equivalent widths of individual 
lines is difficult. This is the reason for invoking the simple analytical relation 
of $v_{\rm t}$ as function of $T_{\rm eff}$ (and $\log g$). 
} ambiguities amounting up to a few tens of \% would be likely, which may lead to 
considerable uncertainties in the absolute abundance of a given star (especially at 
late A-type) by $\lesssim$~0.4~dex, though their impact on relative abundances
between stars would be less critical.
\item
The abundance sensitivity to changing $T_{\rm eff}$ by $\pm 3$\% depends upon $T_{\rm eff}$.
That is, while it is comparatively insignificant in G-type stars ($\sim 0.1$~dex),
its effect becomes progressively important with an increase in $T_{\rm eff}$ 
(up to $\sim$~0.2--0.3~dex) in the regime of A-type stars.
\item
The effect of changing $\log g$ upon the Ba abundance is so small and practically
negligible.
\end{itemize}

\section{Discussion on the trends of barium abundances}

\subsection{Final results of the combined sample}

Now that the non-LTE barium abundances have been determined from
Ba~{\sc ii} 6141/6496 lines for Hyades A and F--G stars, we can discuss
their trends in context of the intended aims described in Sect.~1.
Since $A_{6141}^{\rm N}$ and $A_{6496}^{\rm N}$ are consistent with 
each other (cf. Sect~6.1 and Fig.~7o; note that such a similarity is 
observed also in $A^{\rm L}$ as well as in $\Delta$), 
we adopt the simple average of these two as the final 
$A^{\rm N}$(Ba) to be used for discussion.

The resulting $A^{\rm N}$(Ba) for each of the 89 Hyades stars are plotted 
against $T_{\rm eff}$, $v_{\rm e}\sin i$, and $A$(Fe) in Fig.~8a, 8c, and
8d, respectively, where the supplementarily derived $A_{6141}^{\rm N}$ values 
of 80 field A-type stars (cf. Appendix~A) are also overplotted for reference. 
Abundance trend at each spectral type (G, F, and A) is separately discussed 
based on these figures in the following subsections.

\setcounter{figure}{7}
\begin{figure}
\begin{minipage}{80mm}
\includegraphics[width=8.0cm]{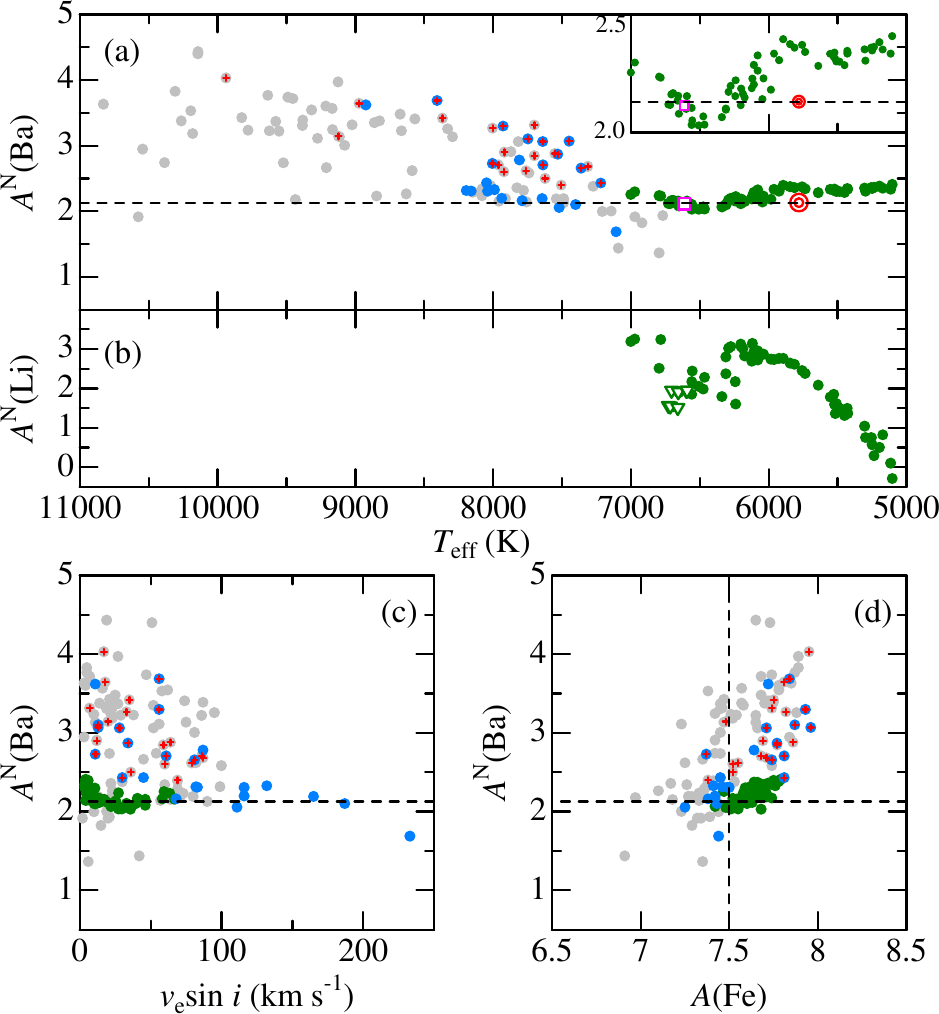}
\caption{
(a) Final non-LTE Ba abundances (mean of $A_{6141}^{\rm N}$ and $A_{6496}^{\rm N}$) 
are plotted against $T_{\rm eff}$ for 23 Hyades A stars and 66 Hyades F--G stars
as well as Sun/Procyon by using the same symbols as in Fig.~7. In addition, 
$A_{6141}^{\rm N}$ results of 80 field A stars (cf. Appendix A)
are also shown with gray symbols for comparison. A-type stars classified as
metallic-line stars (Am stars) are indicated by overplotting red crosses (+).
In the inset are also shown the data of F--G type stars 
($7000 \ge T_{\rm eff} > 5000$~K) in the magnified scale of the ordinate
($2.0 \le A^{\rm N} \le 2.5$).
(b) Non-LTE Li abundances of Hyades F--G stars derived by Takeda et al. (2013)
are plotted against $T_{\rm eff}$ (downward open triangles are upper limit values 
for non-detection cases).
(c) Ba abundances plotted against $v_{\rm e}\sin i$.
(d) Ba abundances plotted against $A$(Fe). 
Dashed lines indicate the positions of reference solar abundances: 
$A_{\odot}$(Fe) = 7.50 and $A_{\odot}$(Ba) = 2.13.
}
\label{fig8}
\end{minipage}
\end{figure}

\subsection{G stars: homogeneous composition}

As recognized from Fig.~8a, $A^{\rm N}$(Ba) values for Hyades G stars 
($6000 \gtrsim T_{\rm eff} \gtrsim 5000$~K) are almost homogeneous without showing 
any systematic dependence upon $T_{\rm eff}$. The average abundance for the relevant 
28 stars is $\langle A^{\rm N} \rangle = 2.33$ (with the standard deviation of 
$\sigma = 0.04$). Since this value may be regarded as the primordial barium abundance 
of the Hyades cluster, it is supersolar by $\sim +0.2$~dex in comparison with 
$A_{\odot} = 2.13$, which is reasonable in view of the mildly metal-rich nature of Hyades 
($0.1 \lesssim {\rm [Fe/H]} \lesssim 0.2$; cf. Fig.~32.8a in Takeda 2008). 

\subsection{F stars: indication of Ba-dip}

Regarding Hyades F stars ($7000 \gtrsim T_{\rm eff} \gtrsim 6000$~K), 
it was unexpected that (unlike G-type stars) their Ba abundances are 
not uniform but show a systematic $T_{\rm eff}$-dependent trend. That is, 
$A^{\rm N}$ decreases with an increase in $T_{\rm eff}$ at $\sim$~6000--6500~K,
and then turns to increase at $\sim$~6500--7000~K, resulting a broad depression 
around $T_{\rm eff} \sim 6500$~K (cf. Fig.~8a). The depth of this dent is
$\sim 0.3$~dex, as estimated from $A^{\rm N}\sim $~2.0--2.1 (bottom at $\sim 6500$~K) 
and $A^{\rm N}\sim $~2.3--2.4 (outside at $\sim 6000$~K). 
 
Similar Ba deficiency in Hyades F stars at $6800 \gtrsim T_{\rm eff} \gtrsim 6400$~K 
is also recognized in Fig.~8 of Varenne \& Monier (1999), who carried out chemical 
abundance determinations for Hyades A--F stars, though they did not mention 
anything about it.
 
Interestingly, the $T_{\rm eff}$ range of this Ba depression almost coincides 
with that of the well-known ``Li dip'' (cf. Fig.~8b) first reported by 
Boesgaard \& Tripicco (1986), which makes us suspect that both might be somehow 
associated with each other. 

However, the physical process causing the Li dip is not yet clarified in spite of 
various attempts from the theoretical side (e.g., Vick et al. 2010; Michaud et al. 2015).
Besides, observational trials of searching for such a specific depression in the abundances 
of other light elements for Hyades F stars have failed (Garc\'{\i}a-L\'{o}pez et al. 
1993 for O; Takeda et al. 1998 for N and O; Takeda et al. 2013 for C and O),
except that only Be ($Z = 5$) may show somewhat similar trend (though much less 
in extent and not clearly) (cf. Boesgaard \& Budge 1989; Boesgaard \& King 2002). 
 
Therefore, although the apparent similarity is surely intriguing, it may be premature 
to consider that a common physical mechanism could be involved in these two phenomena 
concerning light ($Z = 3$) and heavy ($Z = 56$) elements. Further Ba abundance studies 
for F stars in other open clusters of various ages would be necessary and worth trying.

\subsection{A stars: diversified Ba excess}

An inspection of Fig.~8 reveals the characteristic behaviors of Ba abundances
in A-type stars ($T_{\rm eff} \gtrsim 7000$~K) as summarized below, where the 
trends of [Ba/H] ($\equiv A^{\rm N} - 2.13$) for the combined sample of 23 
Hyades A and 80 field A stars are described.
\begin{itemize}
\item
As seen from Fig.~8a, a tendency of increasing [Ba/H] is observed with an  
increase in $T_{\rm eff}$; i.e., from [Ba/H]~$\sim 0$ (at $T_{\rm eff} \sim 7000$~K) 
to [Ba/H]~$\lesssim 2$ (at $T_{\rm eff} \sim 10000$~K).
That is, while Ba anomaly is absent in F-type stars, it begins to appear 
at late A ($T_{\rm eff} \gtrsim 7000$~K) and this Ba-excess tends to be 
more enhanced towards early A ($T_{\rm eff} \sim 10000$~K). 
\item
However, an appreciable dispersion of [Ba/H] is observed for a given $T_{\rm eff}$,
(as seen in Fig.~8a for Hyades stars at $8000 \gtrsim T_{\rm eff} \gtrsim 7000$~K), 
which implies that another parameter (apart from $T_{\rm eff}$) affecting [Ba/H] also exists.
This second parameter should be $v_{\rm e}\sin i$, since the dispersion of 
[Ba/H] tends to  shrink with an increase in $v_{\rm e}\sin i$ 
(from [Ba/H]~$\sim$~0--2 at $\sim 0$~km~s$^{-1}$ down to [Ba/H]~$\sim 0$ at 
$\sim 100$~km~s$^{-1}$) and [Ba/H]~$\sim 0$ holds in the rapid rotator regime 
($100 \lesssim v_{\rm e}\sin i \lesssim 200$~km~s$^{-1}$) (Fig.~8c).
\item
Accordingly, we may regard that [Ba/H] values of A-type stars are controlled 
by $T_{\rm eff}$ and $v_{\rm e}\sin i$. But the latter is more essential
because Ba anomaly ([Ba/H] $> 0$) takes place only for comparatively slower
rotators ($v_{\rm e}\sin i \lesssim 100$~km~s$^{-1}$), while [Ba/H]~$\sim 0$ 
holds for rapid rotators ($v_{\rm e}\sin i \gtrsim 100$~km~s$^{-1}$) 
irrespective of $T_{\rm eff}$.\footnote{Strictly speaking, this 
argument is confirmed from our data only for late A-type stars 
($8000 \gtrsim T_{\rm eff} \gtrsim 7000$~K). But much can not be said for 
early A-type stars ($T_{\rm eff} \gtrsim 8000$~K) because most of them 
have comparatively lower rotational velocities 
(note that $v_{\rm e}\sin i$ values of  all field A stars are lower than 
100~km~s$^{-1}$; cf. Appendix~A) and few rapid rotators of  
$v_{\rm e}\sin i \gtrsim 100$~km~s$^{-1}$ are included.}  
Meanwhile, the difference in $T_{\rm eff}$ should also contribute to the large dispersion 
of [Ba/H] for stars of $v_{\rm e}\sin i \lesssim 100$~km~s$^{-1}$, in the sense 
that higher $T_{\rm eff}$ stars tend to show larger [Ba/H].
\item
A-type stars classified as Am (metallic-line star) tend to have comparatively larger 
Ba excess than those classified as normal (non-Am), as can be recognized from Fig.~8.
This does not mean, however, that non-Am stars have normal Ba abundances.
As a matter of fact, Ba anomaly is observed irrespective of the assigned spectral 
type, so that simple dichotomy into ``normal'' and ``peculiar'' is not much meaningful.  
\end{itemize} 

One of the motivations of this study was to see whether the problem of considerably 
subsolar [Ba/H] seen in A-type stars of near-normal metallicity resulting from previous 
LTE studies (Takeda et al. 2008, 2009; Varenne \& Monier 1999) could be settled by
taking into account of the non-LTE effect. Unfortunately, it was unable to give a
decisive answer based on this analysis alone. 

Regarding Hyades A stars, since most of them are late-A type (mostly $T_{\rm eff} \lesssim 8500~K$), 
differences between $A^{\rm L}$ and $A^{\rm N}$ are not large because non-LTE corrections 
are small at this regime of comparatively lower $T_{\rm eff}$. 
For example, seven Hyades A stars of $v_{\rm e}\sin i > 100$~km~s$^{-1}$ (HD~028024, 027946, 030780, 
029488, 028527, 032301, 028910) have near-solar metallicities ($\langle A({\rm Fe})\rangle = 7.41$).
Their averaged (LTE and NLTE) Ba abundances are $\langle A^{\rm L} \rangle = 2.02$ 
([Ba/H]$^{\rm L} = -0.11$) and $\langle A^{\rm N} \rangle = 2.12$ ([Ba/H]$^{\rm N} = -0.01$);
i.e., the difference is only marginal by 0.1~dex.

As for field A-type stars studied in Appendix A, they also show a correlation between [Ba/H] and 
[Fe/H] (d[Ba/H]/d[Fe/H]~$\sim 4$) but the intersection of this relation at [Fe/H]~$\sim 0$ is
appreciably ``positive'' (not negative!) as   
$0.5 \lesssim {\rm [Ba/H]}^{\rm L} \lesssim 1$ (LTE) and 
$0.5 \lesssim {\rm [Ba/H]}^{\rm N} \lesssim 1.5$ (NLTE; cf. gray symbols in Fig.~8d).  
That is, unlike previous studies mentioned above, the problem of considerably subsolar tendency 
of [Ba/H]$^{\rm L}$ at [Fe/H]~$\sim 0$ does not exist in the first place for these field stars 
(taken from Takeda et al. 2018), reflecting the difference of adopted samples.
Therefore, what can be said based on the present results is that [Ba/H] can be more or less
overabundant in field A-type stars of [Fe/H]~$\sim 0$ (not necessarily [Ba/H]~$\sim 0$ as first
speculated).

\section{Summary and conclusion}

Photospheric abundances of barium in A-type stars (especially sharp-lined 
ones) are known to suffer characteristic peculiarities of overabundance 
(possibly in correlation with [Fe/H] and $v_{\rm e}\sin i$).   
 
However, the quantitative extent of Ba anomaly (acquired abundance change
compared to the initial composition) is still poorly understood.  
For example, according to the past work on Ba abundances of A stars 
(e.g., Takeda et al. 2008, 2009), while $A$(Ba) is correlated with $A$(Fe), 
it is puzzling that stars of near-normal metallicity ([Fe/H]~$\sim 0$) have 
appreciably negative [Ba/H] values by a few tens dex. 
Since this may stem from the assumption of LTE adopted in those 
previous studies, it is desirable to take into account the non-LTE effect.
Besides, for the purpose of assessing the extent of acquired peculiarity, 
it is desirable to establish the original Ba abundance at the time of star formation, 
for which the best approach would be to study the abundances of stars (other than 
A-type) belonging to the same cluster.

Motivated by this consideration, with an aim of clarifying the nature of compositional 
Ba anomaly in A-type stars, spectroscopic determination of Ba abundances based on 
Ba~{\sc ii} 6141/6496 lines were carried out for 89 stars of A-, F-, and G-type, 
($9000 \gtrsim T_{\rm eff} \gtrsim 5000$~K)  belonging to the Hyades cluster, 
while taking into account the non-LTE effect and the hyper-fine-structure effect 
(also significant for Ba~{\sc ii} lines). 

Our statistical-equilibrium calculations for Ba~{\sc ii} carried out for
a grid of models atmospheres covering wide parameter ranges revealed that, 
while the non-LTE effect tends to strengthen lines in G stars (negative non-LTE 
correction), it acts in the direction of line weakening (positive non-LTE 
correction) in the regime of A stars because of the increasing importance of 
overionization.

The Ba abundances of Hyades G stars ($6000 \gtrsim T_{\rm eff} \gtrsim 5000$~K) 
turned out fairly uniform with the mean abundance of  $\langle A \rangle$ = 2.33 
(standard deviation is $\sigma = 0.04$), This indicates that the primordial Ba 
abundance of Hyades is higher by $\sim$~0.2~dex than the solar abundance (2.13), 
which is reasonable in context with the moderately metal-rich nature of the 
Hyades cluster stars. 

A-type stars at $T_{\rm eff} \gtrsim 7000$~K (where 80 field stars were also included 
in addition to 23 Hyades stars) turned out to be generally overabundant in Ba 
with considerably large dispersion (0~$\lesssim$~[Ba/H]~$\lesssim 2$).
The extent of this Ba-excess tends to be more enhanced with an increase in $T_{\rm eff}$,
while the dispersion of [Ba/H] quickly shrinks with an increase of $v_{\rm e}\sin i$
(from $\sim 0$~km~s$^{-1}$ to $\sim 100$~km~s$^{-1}$) and [Ba/H]~$\sim 0$ almost 
holds at $v_{\rm e}\sin i \gtrsim 100$~km~s$^{-1}$. 
This means that the extent of Ba anomaly in A stars is essentially controlled 
by $v_{\rm e}\sin i$ and $T_{\rm eff}$.

Our analysis could not give a decisive answer to the problem of appreciably underabundant 
[Ba/H] in normal A-type stars seen in previous LTE studies (one of the aims of this study), 
because our sample stars do not show such a tendency. 
Yet, since the non-LTE correction is positive (by up to several tenths dex) in the regime 
of A stars, inclusion of this non-LTE effect should act in the direction of increasing [Ba/H] 
(i.e., mitigating the extent of subsolar tendency in the LTE case).   

Finally, regarding Hyades F-type stars ($7000 \gtrsim T_{\rm eff} \gtrsim 6000$~K), 
a rather unexpected result was found that their Ba abundances are not uniform but 
show a broad depression by $\lesssim 0.3$~dex around $T_{\rm eff} \sim 6500$~K.
It is interesting that the $T_{\rm eff}$ range of this Ba-dent almost coincides  
with that of the well-known Li-dip. Whether these two similar phenomena
are actually related with each other is yet to be further investigated. 

\section*{Acknowledgements}

This investigation has made use the VALD database operated at Uppsala University,
the Institute of Astronomy RAS in Moskow, and the University of Vienna.

\section*{Data availability}

The basic data and the results of this investigation are presented as 
the online supplementary material. 

\section*{Supporting information}

This article accompanies the following online materials.
\begin{itemize}
\item
ReadMe.txt 
\item
hyadesA.dat 
\item
hyadesFG.dat 
\item
fieldA.dat 
\end{itemize}

\appendix

\section{Ba abudances of field A-type stars}

A shortcoming concerning our basic program stars of Hyades members is 
that most of the 23 A stars are of late A-type while early A-type stars 
of higher $T_{\rm eff}$ are apparently insufficient (7 at $T_{\rm eff} > 8000$~K, 
only 1 at $T_{\rm eff} > 8500$~K, and none at $T_{\rm eff} > 9000$~K; cf. Fig.~2). 
It was necessary, therefore, to increase the number of A stars in our 
sample by invoking field stars.

Takeda et al. (2018) once conducted CNO abundance determinations for 100 A-type 
main-sequence stars of comparatively slow rotators ($v_{\rm e}\sin i < 100$~km~s$^{-1}$) 
based on the high-dispersion spectra obtained at OAO/HIDES and BOAO/BOES.
Accordingly, since 20 stars (among 100 stars) are already common to our 23 Hyades 
A stars, Ba abundances of the remaining 80 stars were newly determined by using 
the same observational data and the same atmospheric parameters/models adopted therein. 
The analysis was done in the same manner as described in Sect.~4 and Sect.~6, but 
only the Ba~{\sc ii} 6141 line was employed in this supplementary work.

The atmospheric parameters ($T_{\rm eff}$, $\log g$, $v_{\rm t}$, $v_{\rm e}\sin i$ 
and $A$(Fe)) as well as the resulting abundance-related quantities ($W_{6141}$, 
$A^{\rm L}_{6141}$, $\Delta_{6141}$, and $A^{\rm N}_{6141}$) for these 80 field A 
stars are presented in ``fieldA.dat'' of the online material, and they are 
graphically displayed in Fig.~A1 (arranged in the same format as in Fig.~2a--2d
and Fig.~7a--7d). These data (overplotted in Fig.~8) are employed for discussion 
in Sect.~7.4 in combination with those of Hyades stars.

\setcounter{figure}{0}
\begin{figure}
\begin{minipage}{80mm}
\includegraphics[width=8.0cm]{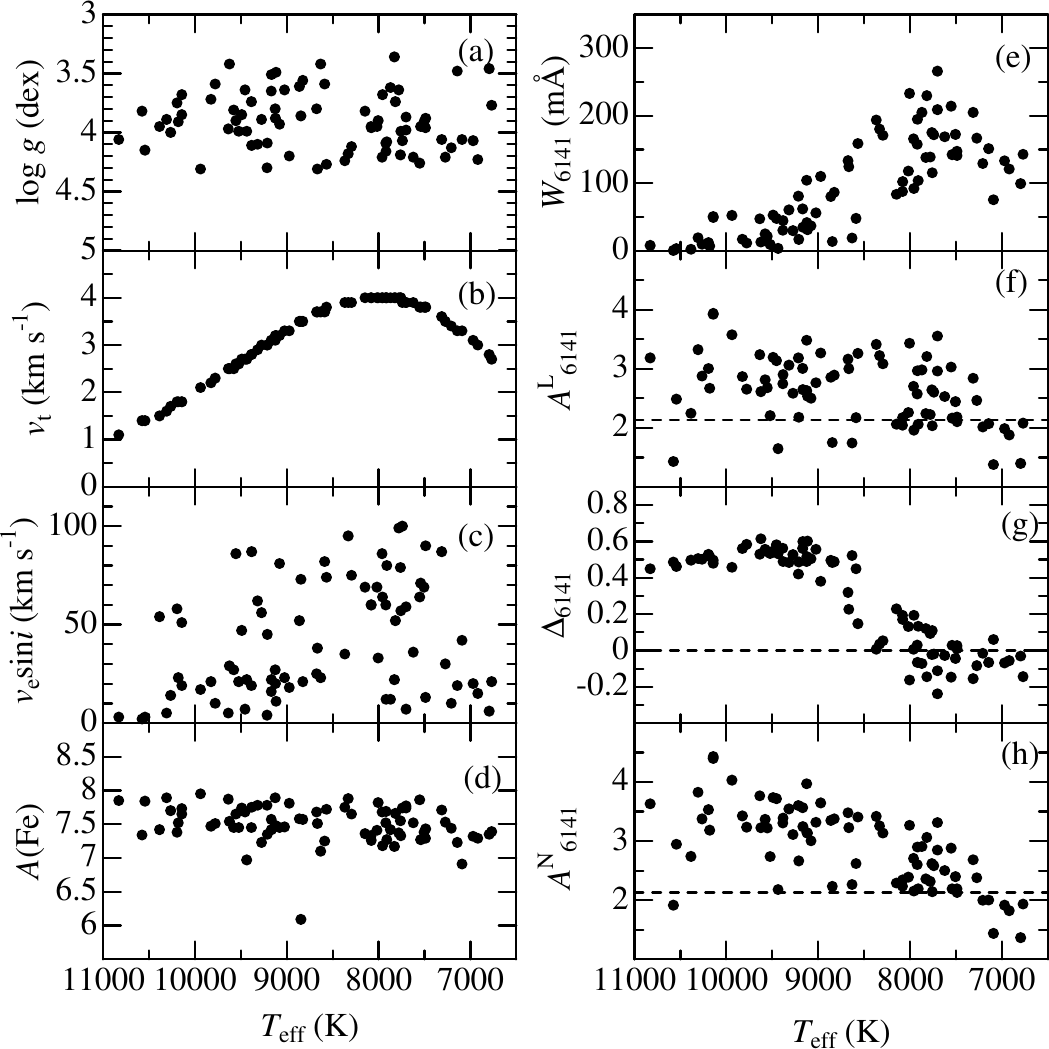}
\caption{
Left panels (a)--(d): atmospheric parameters of 80 field A-type stars
($\log g$, $v_{\rm t}$, $v_{\rm e}\sin i$, and $A$(Fe)) are plotted 
against $T_{\rm eff}$, as done in Fig.~2a--2d for Hyades stars.
Right panels (e)--(h): Ba~{\sc ii} 6141 line-related quantities 
($W_{6141}$, $A_{6141}^{\rm L}$, $\Delta_{6141}$, and $A_{6141}^{\rm L}$)
derived for 80 field A stars are plotted against $T_{\rm eff}$, as done
in Fig.~7a--7d for Hyades stars.
}
\label{figA1}
\end{minipage}
\end{figure}

\end{document}